\documentclass[prd,twocolumn]{revtex4}%
\usepackage{amsmath}
\usepackage{graphicx}
\usepackage{epsfig}
\usepackage{dcolumn}
\usepackage{bm}
\usepackage{dcolumn}
\usepackage{amsfonts}
\usepackage{amssymb}%
\usepackage{orcidlink}
\usepackage{float}
\usepackage{hyperref}
\providecommand{\U}[1]{\protect\rule{.1in}{.1in}}
\begin{document}
\title{About the Upstream Contamination\footnote{This article has been published in the Revista Cubana de Física, Vol. 36, No. 1, pp. 8-14, 2019.} }
\author{D. Suárez-Fontanella$^{123}$ , A. Cabo Montes de Oca $^{1}$ }
\affiliation{$^{1}$Departamento de F\'{\i}sica Te\'{o}rica, Instituto de
Cibern\'{e}tica Matem\'{a}tematica y F\'{\i}sica (ICIMAF), Calle E, No. 309,
entre 13 y 15, Vedado, La Habana, Cuba.}
\affiliation{$^{2}$Facultad de F\'{\i}sica de la Universidad de la Habana, La Habana, CP 10400, Cuba\\ $^{3}$ Facultad de Física Fundamental y Matemáticas Universidad de Salamanca, Plaza de la Merced S/N, Salamanca, E-37008, Spain }

\begin{abstract}
This study investigates the upward transport of waterborne pollutants from a lower container to an upper container through vertically falling water streams. While previous analyses have primarily focused on inclined channels, we extend the theoretical framework to consider vertical configurations. Two distinct cases are examined: (i) a vertical flow within a cylindrical tube, and (ii) a free-falling water jet.
For the first case, we derive an analytical expression for the critical water flux required to prevent the upward migration of particles. In the second case, we establish a relationship among water flux, particle size, and vertical position along the stream, which determines the feasibility of upward particle transport.
Our findings reveal a fundamental difference between the two configurations. In the tubular flow case, surface tension has negligible influence on particle motion. In contrast, in the free-fall scenario, upward particle transport is only possible in the presence of surface tension. Moreover, we demonstrate that for any given water flux, there exists a threshold height difference beyond which contamination of the upper container is not possible. Increasing the water flux further inhibits any upward transport of pollutants.
 \bigskip

\noindent D. Suárez-Fontanella E-Mail: \href{mailto:duvier@usal.es}{duvier@usal.es}

\noindent A. Cabo E-Mail: \href{mailto:cabo@icimaf.cu}{cabo@icimaf.cu}

\end{abstract}

\maketitle

\section{Introduction}

The present work is devoted to discussing the so-called \textit{upstream contamination} effect. This phenomenon was detected when a jet of water was falling into a container where \textit{mate} particles were floating. It was then noticed that after some time, the particles also appeared in the higher vessel.

This property was observed in 2008 by a student from the Faculty of Physics at Havana University and was reported and further investigated in reference \cite{Bianchini}. In that work, the analysis mainly concentrated on explaining the experimental setup developed for this study. It consisted of two reservoirs lying at different heights, connected by an inclined rectangular channel. At the lower end of the channel, there was also a free-falling stream of water.\\

The experimental observations carried out in that work were mainly explained by invoking the so-called Marangoni effect \cite{Reynolds, Marangoni, Levich, Screvin, Singh}, which consists of the variation of the surface tension of substances when floating contaminants are present. Then, the gradient of the local density of surface contaminants was assumed to create a vertical floating force \cite{Bianchini}.

In this work, we will present a complementary discussion of the contamination effect, limited to the simpler case of vertically falling water beams. The objective is to consider the feasibility of the effect occurring without the additional influence of the Marangoni effect. That is, only the effects of viscosity and surface tension will be considered. Specifically, we will determine conditions on the parameters that must be satisfied for spherical particles to ascend through the water jet. The discussion will assume that the particles move through the interior of the water beam. The case in which the particles are elevated through the surface will not be analyzed here.

Two different situations will be investigated. In the first case, the water will fall through a cylindrical tube with a circular cross-section. In the second case, the water is assumed to fall freely under gravity. In both cases, the Stokes frictional force associated with water viscosity plays a central role. However, an interesting outcome is that while surface tension plays no role in the first case, it becomes central in the second. For example, it follows that in the absence of surface tension, the particles are unable to ascend. In other words, the floating force that makes the particles ascend is entirely determined by the surface tension.

The results of the first case determine the flux of water over which the particles are stopped from flowing up. For the second system discussed, it follows that for every value of the particle size and water flux, there is a critical height difference between the recipients over which the particles turn out to be unable to flow up. This critical value grows with the increase in the size of the particles and with the reduction in the water flux. It follows that the numerical values predicted by the obtained relations for the critical height, water fluxes, and particle sizes of the free-falling water beam at the end of the channel are in the range of the experimentally observed values in reference \cite{Bianchini}.

The presentation of the work will be as follows. In Section 2, the relations among the parameters for stopping the water from contaminating the higher recipient in the simpler case of water falling through a vertical tube will be discussed. Section 3 considers the derivation of the analogous relation for the case of the free-falling water beam. The determination of an approximate stationary form of the beam is also presented. Finally, the results are reviewed in the Summary.

\section{Water falling through a cylindrical tube}

Let us assume that the reference frame for coordinates will be at the bottom of the upper recipient, so that the positive axis of the $z$ coordinate points downwards and coincides with the axis of symmetry of the water flow. Under these assumptions, the Bernoulli theorem applied between any two points $a$ and $b$, lying at different heights and both lying on a curve tangent to the velocity field at any point, can be written in the form:

\begin{equation}
\frac{1}{2}\rho\text{ }v_{a}^{2}-\rho\text{ }g\text{ }z_{a}+P_{a}=\frac{1}%
{2}\rho\text{ }v_{b}{}^{2}-\rho\text{ }g\text{ }z_{b}+P_{b}%
.\ \ \ \ \ \label{bernoulli}%
\end{equation}
In the following, this relation will be used to discuss the two types of water flows under consideration. Given that the index $a=0$ and the index $b$ is defined by the $z$ axis coordinate, the Bernoulli law can be written as:
\begin{align*}
\frac{1}{2}\rho v_{0}^{2}+P_{0}  &  =\frac{1}{2}\rho\text{ }v(z)^{2}-\rho
g\text{ }z+P(z),\\
\frac{1}{2}\rho v_{0}^{2}  &  =\frac{1}{2}\rho\text{ }v(z)^{2}-\rho\text{
}g\text{ }z+(P(z)-P_{0}).
\end{align*}

However, for the case under consideration in this section, the cross-sectional area of the cylinder remains constant at different heights. Therefore, the incompressibility of water implies that the velocity \( v(z) \) does not change with \( z \), and

\begin{equation}
v(z)=v_{0}\equiv v.
\end{equation}
Thus, it also follows that the distribution of pressures along the
vertical inside the tube is identical to the one in static water, thus
\begin{equation}
P(z)-P_{0}=\rho\text{ }g\text{ }z.
\end{equation}
Now, let us write the Newton equation of motion for a small spherical particle of
radius $R$ and density $\rho_{m}$ as
\begin{align}
\rho_{m}\text{ }V\text{ }\frac{dv_{m}(t)}{dt} &  =\text{ }\rho_{m}V\text{
}g+f_{e}-k\text{ }(v_{m}(t)-v),\label{newt}\\
m &  =\rho_{m}\text{ }V,\\
V &  =\frac{4}{3}\pi R^{3},
\end{align}
This expression shows that the acceleration is determined by the vector sum of the body's weight, the buoyant forces due to pressure, and the viscous force defined by Stokes' law. However, Archimedes' law is exactly valid for the buoyant force due to the linear pressure behavior, which coincides with the functional dependence of pressure on vertical distance in static water. Thus, the buoyant force is given by:

\begin{equation}
f_{e}=-V\frac{dP(z)}{dz}=-\rho\text{ }g\text{ }V,
\end{equation}
where the negative sign is because the force tend to move the body in the
negative sense of the height coordinate $z$. Then, the Newton equation
 can be rewritten in the form
\begin{equation}
\frac{dv_{m}(t)}{dt}=\text{ }g\text{ }(1-\frac{\rho\text{ }}{\rho_{m}}%
)-\frac{k\text{ }}{m}(v_{m}(t)-v).
\end{equation}

This  equation  includes the frictional Stokes force acting on a
moving sphere in a fluid when the liquid movement is laminar
\begin{equation}
f_{S}=-\text{ }k\text{ }(v_{m}(t)-v),
\end{equation}
where the constant $k$ is given in terms of the viscosity constant $\mu$ and
the radius of the sphere as
\begin{equation}
k=6\text{ }\pi\mu\text{ }R.
\end{equation}

This is a simple first order equation for the velocity of the particle in the
observer's reference frame. Expressing it in a integral form, \ we can write
\begin{align}
\int_{0}^{v(t)}\frac{dv_{m}}{(v_{m}(t)-v)+\frac{m\text{ }g}{k}\text{ }%
(\frac{\rho\text{ }}{\rho_{m}}-1)}  &  =-\int_{0}^{t}\frac{k}{m}dt,\nonumber\\
\log{\large [}\frac{v_{m}(t)-v+\frac{m\text{ }g}{k}\text{ }(\frac{\rho\text{
}}{\rho_{m}}-1)}{-v+\frac{m\text{ }g}{k}\text{ }(\frac{\rho\text{ }}{\rho_{m}%
}-1)}{\large ]}  &  =-\frac{k}{m}t.
\end{align}
This relation allow to write the following explicit solution for the  velocity
of the particle%
\begin{equation}
v_{m}(t)=[v-\frac{m\text{ }g}{k}\text{ }(\frac{\rho\text{ }}{\rho_{m}%
}-1)](1-\exp(-\frac{k}{m}t)),
\end{equation}
which after evaluated at large times determines the following expression for  the  limit velocity of the
particle with respect to the observer's frame
\begin{align}
v_{m}(\infty)  &  =[v-\frac{m\text{ }g}{k}\text{ }(\frac{\rho\text{ }}%
{\rho_{m}}-1)]\\
&  =v-\frac{2\pi R^{2}g\text{ }\rho_{m}}{9\text{ }\pi\mu\text{ }}\text{
}(\frac{\rho\text{ }}{\rho_{m}}-1).\nonumber
\end{align}
In the above relations  it has been substituted
\begin{equation}
\frac{m}{k}=\frac{2\pi R^{2}\rho_{m}}{9\text{ }\pi\mu\text{ }}.
\end{equation}
Therefore, the single condition for the particles not to be allowed to climb
to the upper reservoir is the satisfaction of the inequality%
\[
v-\frac{2\pi R^{2}g}{9\text{ }\pi\mu\text{ }}\text{ }(\rho-\rho_{m})>0.
\]

But, the velocity is defined in terms of the volume flux of water $Q$ and the
area $A$ of the vertical cylinder as $v=\frac{Q}{A}$. Thus, the inequality
becomes fully expressed in terms of the parameter of the system in the form
\begin{equation}
Q>\frac{2\pi R^{2}gA}{9\text{ }\pi\mu\text{ }}\text{ }(\rho\text{ }-\rho_{m}).
\end{equation}

This relation indicates that the if flux will stop all the particles whatever
the value of their density $\rho_{m}$ the following modified relation \ should
be satisfied
\begin{equation}
Q>\frac{2\pi R^{2}gA\rho}{9\text{ }\pi\mu\text{ }}\text{ .}\label{cond2}%
\end{equation}

\begin{figure}[ht]
\centering
\includegraphics[width=8.5cm]{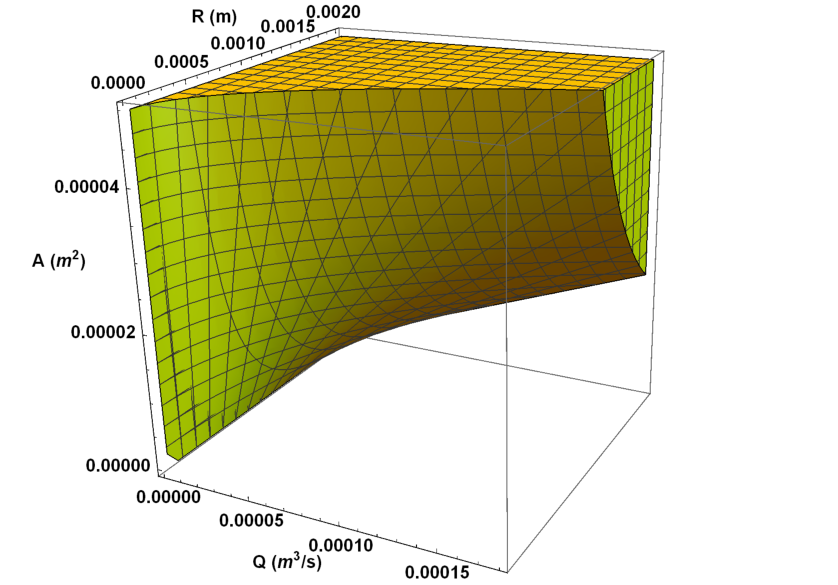}
\caption{ The figure illustrates a region in the parameter space defined by the water flux $Q$, the radial size $R$ of the pollutant particles, and the cross-sectional area $A$ of the cylindrical tube. The plotted surface corresponds to the set of points that satisfy the equality given in equation~(\ref{cond2}). At these points, the velocity of the pollutant particles becomes zero when the parameter triplet $(Q, R, A)$ takes the corresponding values. Below this surface, the particles acquire a net downward velocity and therefore do not contaminate the upper container. In contrast, above the surface—in the colored region—the particles tend to ascend against the flow.
}%
\label{tubo}%
\end{figure}

Let us now assume that the fluid is water and the experiment is conducted under normal gravity conditions. Then, the following parameters are used: \( \rho = 1000\ \text{Kg/m}^{3}, g = 9.8\ \text{m/s}^{2}, \mu = 8.9 \times 10^{-4}\ \text{Ns/m}^{2}. \) We now illustrate the resulting regions of the water flux \( Q \), the radius \( R \) of the spherical particles, and the area \( A \) of the water-conducting cylinder, for which the particles are either allowed or not allowed to ascend through the cylinder. These regions are shown in figure \ref{tubo}. The surface shown defines the critical boundary in parameter space, where the particles are on the threshold of either being allowed or not allowed to contaminate the upper recipient. The colored zone above the surface represents the set of values of the triplet \( (Q, R, A) \) for which the particles can move against the flow towards the higher recipient. Conversely, the white region below the surface represents the triplets of parameters for which the particles remain trapped in the lower vessel.

Let us use the derived relation above to estimate the critical value of the flux for particles of chalk powder, as used in reference \cite{Bianchini}, which serve as pollutant particles. In this work, the sizes of these particles were estimated to be on the order of hundreds of micrometers ($\mu$m). Thus, we will assume the radius of the spherical particles to be \( R = 100 \ \mu\text{m} = 0.0001 \ \text{m} \). The water flux falling from the upper to lower recipients is chosen as \( Q = 16 \ \text{cm}^{3}/\text{s} = 0.000016 \ \text{m}^{3}/\text{s} \). For these values of particle size and water flux, the point on the critical surface corresponds to a cylinder sectional area \( A = 0.000163469 \ \text{m}^{2} = 1.63 \ \text{cm}^{2} \). \\

This result indicates that for the water flux used in the experiment in reference \cite{Bianchini}, and the estimated size of the chalk powder particles, the area of the cylinder section considered in our model is on the order of a few cm$^{2}$. The sectional area of the waterfall at the end of the inclined channel used in the experiments is expected to be of the same order. Therefore, it can be concluded that floating effects alone, without considering the Marangoni effect, can at least partially justify the upstream contamination effect through the waterfall at the end of the channel.

We would like to highlight an issue that seems relevant in connection with the experiments reported in reference \cite{Bianchini}. It should be emphasized that the viscous nature of water is expected to result in a zero velocity at the channel walls through which the water flows in the experiment. This situation suggests that the contact boundary between the water and the channel walls may serve as a preferred path for the upward motion of the particles. Therefore, it seems necessary to compare the dimensions of the particles with the size of the boundary region, where the velocity transitions from zero at the wall to the mean values in the center of the channel.

\section{Free falling water flow}

Now, let us consider the second situation in which the water free-falls in a beam to the lower recipient.

\begin{figure}[ht]
    \centering
    \includegraphics[width=0.8\linewidth]{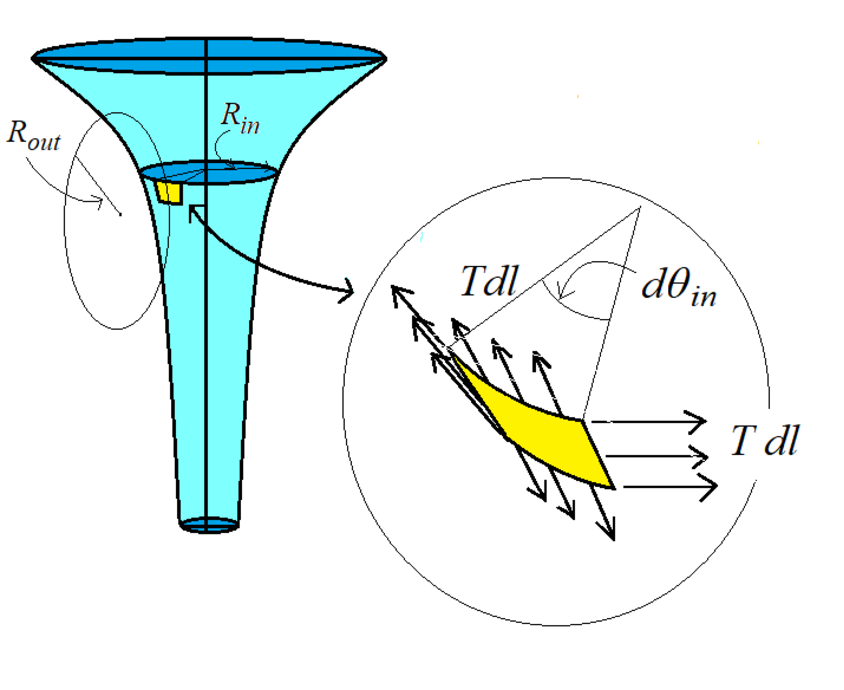}
    \caption{A model of the water beam is presented, incorporating a surface element on which a momentum balance is performed. The internal and external principal curvature radii of the surface are indicated. The surface element is highlighted in yellow, and the surface tension forces acting on it are illustrated in the figure depicted on the right.}
    \label{model}
\end{figure}

First, the pressure difference between two points close to the surface of the water beam will be determined: one point in the air and the other inside the water. Then, consider the momentum balance of a local piece of surface illustrated in Figure \ref{model}. The figure illustrates the two principal curvature radii, $R_{1}$ and $R_{2}$, of the chosen symmetric surface element, and also the surface tension forces primarily exerted on the two arcs of the circle corresponding to the respective circumferences defining the curvature radii.

Therefore, considering that the two arc elements are infinitesimal and of the same size, $dl$, it follows from the balanced momentum equation of the depicted surface elements:

\begin{align}
(P_{in}-P_{out})\text{ }dl\text{ }^{2}  &  =\gamma\text{ }dl\text{ }(d\theta_{in}-d\theta_{out}).\nonumber\\
&  =\gamma\text{ }dl\text{ }^{2}(\frac{1}{R_{in}}-\frac{1}{R_{out}}d\theta
_{out}),\\
d\theta_{in}  &  =\frac{dl}{R_{in}},\text{ \ \ \ }d\theta_{out}=\frac
{dl}{R_{out}}.
\end{align}

Now, in order to simplify the discussion, let us assume  that  water beam shows
 curvature radii  satisfying
\begin{equation}
R_{out}>>R_{in}.
\end{equation}
This relation indicates that  the tangent vector to the beam surface contained in a plane including the axis of the beam,
becomes close in direction  with the beam axis.  But, this property
 in turns implies that the small radius is approximately given by the
radius of the beam taken at the fixed height value defined by the  $z$
coordinate. This radius will be defined by $r(z)$. Thus,
\begin{equation}
P_{in}-P_{out}\text{ }=\frac{\gamma}{r(z)}.
\end{equation}

But, the external pressure to the beam is given by the atmospheric one
$P_{0},$ what leads for the relation between the internal pressure at any
height $z$ the formula%
\begin{equation}
P(z)=P_{0}+\frac{\gamma}{r(z)}.\label{press}%
\end{equation}

Consider now that at the origin of the height coordinate $z=0,$ the
following initial conditions: for the speed of the fluid when emerging form
the upper recipient is $v(0)=v_{0}$ and for the radius of the beam $r(0)=a.$
 Then, substituting in the Bernoulli equation it follows
\begin{align}
\frac{1}{2}\rho v_{0}^{2}+P_{0}+\frac{\gamma}{a}  &  =\frac{1}{2}\rho\text{
}v(z)^{2}-\rho g\text{ }z+P_{0}+\frac{\gamma}{r(z)},\\
\frac{1}{2}\rho v_{0}^{2}+\frac{\gamma}{a}  &  =\frac{1}{2}\rho\text{ }%
v(z)^{2}-\rho\text{ }g\text{ }z+\frac{\gamma}{r(z)}, \label{system}%
\end{align}
from which  the velocity can be expressed in the  form
\[
v(z)^{2}=v_{0}^{2}+2\text{ }g\text{ }z+\frac{2\text{ }\gamma}{\rho\text{ }a}%
-\frac{2\text{ }\gamma}{\rho\text{ }r(z)}.
\]

Let us consider now a  special origin for measuring the height coordinate
$z.$ Note first that if we assume the velocity $v_{0}$ of the flux to tends to
zero in (\ref{system}), the constancy of the total flux of water $Q$ implies
through $Q=\pi a^{2}v_{0}$,  that the radius of the beam $a$ should tend to
infinity. Therefore, if we consider this point as the origin of coordinates
$z=0,$ relation (\ref{system}) becomes
\begin{equation}
\rho\text{ }g\text{ }z=\frac{1}{2}\frac{\rho\text{ }Q^{2}}{\pi\text{ }%
r(z)^{4}}+\frac{\gamma}{r(z)}. \label{inverse}%
\end{equation}

\begin{figure}[ht]
\centering
\includegraphics[width=7.5cm]{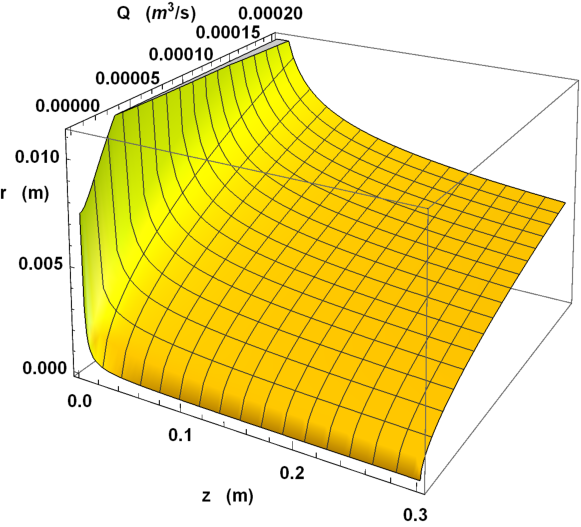}
\caption{ The figure shows the dependence of the radius of the free-falling water beam as a function of the height coordinate $z$ and the water flux $Q$ through the beam. It should be noted that the radius values for small height coordinates $z$ are no longer approximations when the large curvature radius of the beam is not significantly larger than the small curvature radius.
}%
\label{radius}%
\end{figure}

But, the unique real and positive solution of this equation gives for $r(Q,z)$
(after fixing the parameters of the density, surface tension and viscosity
associated to water) is plotted in figure (\ref{radius}) as a function of
the flux and the height $z$ (as measured from the point at which the velocity
of water vanish). Note that this chosen origin of values of $z$ is an
unphysical  point of the curve $r(z)$, since the section of a real  beam never
tends to infinity.

The discussion above in this section describing the flow of water regime in
the falling beam of water mainly follows the one given in reference
\cite{Hancock}. It was presented for completeness reasons.

Let us now consider how a particle situated in the water beam moves. The
forces acting on this particle are  due to:  the variation of the
pressure with the height, the weight of the particles and  the force of viscosity. This last one
 is proportional to the relative speed between the liquid and the
particles (Stokes Law). Then, the Newton Second Law takes the
following form:%
\begin{align}
\rho_{m}\text{ }V\text{ }\frac{d^{2}z}{dt^{2}} &  =\text{ }\rho_{m}V\text{
}g-k\text{ }(\frac{dz}{dt}-v(z))-V\frac{dP(z)}{dz},\label{newton 1}\\
m &  =\rho_{m}\text{ }V,
\end{align}
which is very similar to relation (\ref{newt}) in the past section with the
only difference that the expression for the pressure as a function of the
height $z$ is another one. In this equation $\frac{dz}{dt}=v_{m}(t)$ is the
speed of the particle regarding to the observer and $\rho,V$ (are as in the
past subsection) the density and the volume of the floating particle.
\ But, the derivative of  the pressure in (\ref{press}) can be expressed in terms of  $\frac{d\text{
}r(z)}{dz}$, which in turns  can be expressed as a function of $r(z)$ by
taking  the derivative of the relation (\ref{inverse}).  Using the result of this evaluation,
the floating force (equal to minus the
derivative of the pressure respect to the height $z$ times the volume) can be
calculated in the form
\begin{align}
f_{e} &  =-V\frac{dP(z)}{dz}\nonumber\\
&  =-m\text{ }g\frac{\rho}{\rho_{m}}\frac{\gamma\text{ }r(z)^{3}}{\frac{2\rho\text{
}Q^{2}}{\pi}+\gamma\text{ }r(z)^{3}}.
\end{align}

This result for the floating force deserves a comment. Note that, discarding
the viscosity) the unique force directed in the negative axis of the
coordinate $z$ (that is tending to elevate the particle to the upper
recipient) is this floating force that completely disappears if the surface
tension vanishes. This permits to conclude that,  in the here considered  free
falling case, the surface tension is a central element for the possibility of
contamination of the upper recipient by particles coming from the lower one.

Then, by dividing the Newton equation by the mass of the particle $m=\rho_{m}$
$V$ , it is possible to write
\begin{align}
\frac{d^{2}z}{dt^{2}}&=(  g-\frac{9\pi\mu\text{ }}{2\pi\rho_{m}R^{2}%
}(\frac{dz}{dt}-v(z)) \\ \nonumber &-\text{ }g\frac{\rho}{\rho_{m}}\frac{\gamma\text{ }r(z)^{3}%
}{\frac{2\rho\text{ }Q^{2}}{\pi}+\gamma\text{ }r(z)^{3}})  ,\label{freefall}%
\end{align}
in which all the entering parameters are already well defined. We will
consider now the restrictions implied by this equation on the values these
parameters, allowing or not the transportation of particles from the recipient
down to the upper one. A drastic simplification for the derivation of these
conditions follows after noting that the term in the equation pushing the
particles to go down (tending to positively increase the values of $z$,  which
are positive in the downward direction) is a constant equal to $g$.
Thus, when density $\rho_{m}$ tends to vanish we have the situation in which
there is a maximal tendency of particle to move up. That is the limit in which
the particle is an empty bubble.

In this limit the equation reduces to the simpler form
\begin{equation}
\text{ }0=\text{ }\frac{9\pi\mu\text{ }}{2\pi R^{2}}(\frac{dz}{dt}%
-v(z))+\text{ }g\rho\frac{\gamma\text{ }r(z)^{3}}{\frac{2\rho\text{ }Q^{2}}{\pi
}+\gamma\text{ }r(z)^{3}},
\end{equation}
from which the velocity of the particle at any point can be expressed in terms
of the already known magnitudes of the problem as%
\begin{equation}
\frac{dz}{dt}=v(z)-\frac{2R^{2}g\rho}{9\mu\text{ }}\frac{\gamma\text{ }r(z)^{3}%
}{\frac{2\rho\text{ }Q^{2}}{\pi}+\gamma\text{ }r(z)^{3}}.
\end{equation}

Therefore, the condition for the particle to move up at a height value $z$ can
be written as
\begin{align}
\frac{dz}{dt}  &  =v(z)-\frac{2R^{2}g\rho}{9\mu\text{ }}\frac{\gamma\text{
}r(z)^{3}}{\frac{2\rho\text{ }Q^{2}}{\pi}+\gamma\text{ }r(z)^{3}}\nonumber\\
&  =\frac{Q}{\pi r(z)^{2}}-\frac{2R^{2}g\rho}{9\mu\text{ }}\frac{\gamma\text{
}r(z)^{3}}{\frac{2\rho\text{ }Q^{2}}{\pi}+\gamma\text{ }r(z)^{3}}<0.
\end{align}

Assumed that the fluid under consideration is water, the above conditions is
equivalent to the negative sign of the following function $C$ of the flux $Q$,
radius of the particle $R$ and the height coordinate $z$
\begin{equation}
C(Q,R,z)=Q-\frac{2\pi R^{2}g\rho}{9\mu\text{ }}\frac{\gamma\text{ }r(z)^{5}}%
{\frac{2\rho\text{ }Q^{2}}{\pi}+\gamma\text{ }r(z)^{3}}<0.\label{cond1}%
\end{equation}

\begin{figure}[ht]
\centering
\includegraphics[width=7.5cm]{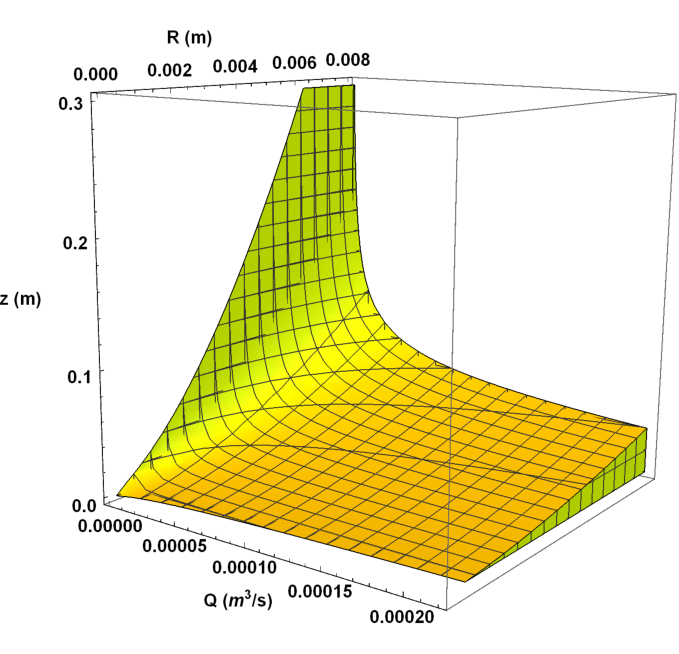}
\caption{ The figure depicts a region in the parameter space defined by the water flux $Q$, the height coordinate $z$ (measured from the critical point where the flow velocity vanishes), and the radial size $R$ of the pollutant particles. The plotted surface corresponds to the set of points satisfying the equality in equation~(\ref{cond1}). That is, at these points, the velocity of the pollutant particles vanishes when the triplet of parameters takes the plotted values.
Above the surface, the particles have positive velocities and therefore do not contaminate the upper container. Below the surface, in the colored region, the particles have negative velocities and thus tend to ascend against the flow.
}%
\label{caidalibre}%
\end{figure}

The satisfaction of the relation (\ref{cond1}) in the space of the three
still free parameters: the flux $Q$, the radius of the particles $R$ and the
height coordinate $z$ (measured from the point of zero velocity) is
illustrated in the figure  \ref{caidalibre}. The shown surface describes the
triplets of values of the parameters $(Q,R,z)$ at which the velocity of the
particle becomes equal to zero. For all the points being over this surface the
particle shows a positive value of its velocity and then it is not able to
climb to the upper reservoir (remember that the positive increments of the
height $z$ are defined as positives if the point moves down). Correspondingly,
the points being below the surface correspond to particles that will
contaminate the upper vessel. That is, having a negative velocity value. In
this way, given the parameters of the system, the conditions for the particle
to appear in the upper recipient have been identified also for this free
falling water situation. In the next subsection we will check how the direct
solution of the exact systems of equations reproduce the same conclusions
extracted from the simplified analysis for a spherical bubble of the same
radius as a particle with density $\rho_{m.}$

\subsection{The direct solution of the full Newton equations near the critical surface.
}

Let us consider the full Newton the systems of equation
\begin{align}
\frac{d^{2}z}{dt^{2}}&=  g-\frac{9\pi\mu}
{2\pi\rho_{m}R^{2}}(\frac{dz}{dt}-v(z))\\\nonumber
&-g\frac{\rho}{\rho_{m}}\frac{\gamma r(z)^{3}}{\frac{2\rho Q^{2}}{\pi}+\gamma r(z)^{3}
}, \label{freefall1}
\end{align}
for values of the triplet of parameters $(Q,R,z)$ being close to the critical surface in figure  \ref{caidalibre}. For concreteness, it will be  assumed the following specific values for the height position and the radius of the particle
\begin{align}
z(0)   =0.1\text{ \ m, } \, R  =0.004\text{ \ m,}
\end{align}
by also selecting two values of the flux
\begin{align}
Q_{1}  = 1.4\times10^{-6}\text{m}^{3}/s  \, \, \, \, \, \,  Q_{2}  = 1.6\times10^{-6}\text{ m}^{3}/s \label{twoQ}
\end{align}
being close to an specific value $Q^{\ast}=1,53$ $\times10^{-6}$. This value $Q^{\ast}$ is chosen for assuring that
 the triplet  $(Q^{\ast},R,z(0))$ is   exactly on the critical surface.  Then, the  two triplets associated to the
  fluxes $Q_{1}$ and $Q_{2}$ are  situated,  one of them over and the other below, the critical surface.

Now,  it will be  considered the solution of the equations  (\ref{freefall1})
 by supposing that the density of the particle is  very small, by example
 satisfying
\[
\frac{\rho_{m}}{\rho}=10^{-9}.
\]
   In addition, we will assume two independent  boundary condition for the velocity of the
    particles as coinciding at the initial time $t=0$ with the velocities of the water flow in the form%
\[
\frac{d}{dt}z(t)_{t=0}=\overset{\cdot}{z}(0)=\frac{Q}{\pi r(z(0))^{2}},
\]
as calculated  for the two specified values of the flux:  one of them
associated to a higher value  and defining  a point over the critical surface
and the other one with a smaller flux, which is linked with a point being
below the critical surface. The specific values of the two fluxes were defined in
(\ref{twoQ}).
\begin{figure}[ht]
\centering
\includegraphics[width=7.5cm]{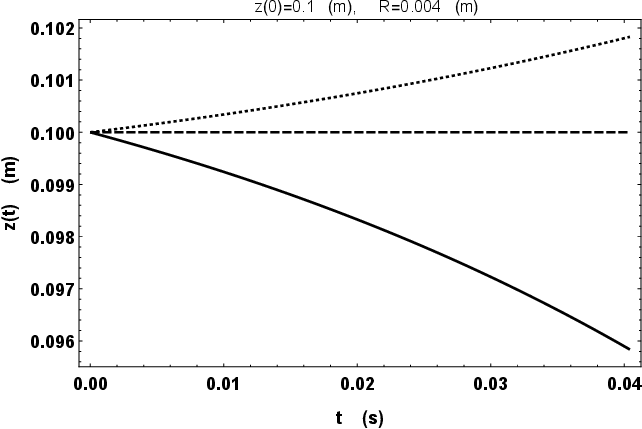}
\caption{ The plotted curves illustrate that when the flux is smaller and the triplet of parameters $(Q, z, R)$ lies below the critical surface, the particle moves towards negative values of $z$. In this case, the particle ascends to the upper reservoir.
On the other hand, for higher flux values (when the triplet of parameters lies above the critical surface), the particle moves towards the positive values of $z$. Thus, in this case, the particle moves downward and does not contaminate the upper vessel.
}%
\label{nearcrit}%
\end{figure}
 The solutions of the equations of motion for the height coordinate $z$ of the particle are shown in Figure \ref{nearcrit}. The curves clearly show that when the flux is smaller, and the set of parameters is below the critical surface, the particle's motion tends to the negative values of $z$. That is, the particle tends to ascend to the upper reservoir. However, for higher flux, when the set of parameters is above the critical surface, the particle tends to move in the direction of the positive values of $z$. Thus, in this case, the particle moves downward and does not tend to contaminate the upper vessel. Then, the direct solutions of Newton's equations intersect the critical surface in the space of parameters defined by solving the exact equations for empty bubbles.

\section{Summary}

We have theoretically modeled the effect associated with the rising of particles from a lower vessel to a higher one, through a water flux falling from the higher vessel \cite{Bianchini}. Conditions for the occurrence of this effect are determined for the set of parameters corresponding to two types of water falling mechanisms. It follows that for water flowing through a cylindrical tube with a constant cross-sectional area, there is a critical flux value above which the particles are not allowed to rise to the upper vessel.

In the case of free falling water, it became clear that the particles can only rise if surface tension is present in the water. If surface forces are assumed to be absent, the particles cannot rise, at least in the discussed case of movement through the volume. Furthermore, in this case, there exists a critical height difference between the vessels above which the particles cannot climb the water flow.

These critical values increase with the particle size and decrease with a reduced water flux through the falling beam. The discussion presents a simple criterion for the occurrence of the effect, which is based on the simplified equation of motion for empty bubbles of the same size as the particles. The almost vanishing of the bubble density allows reducing the equation of motion to first order in the time derivative in the limit of zero density.

The satisfaction of the criterion for the occurrence of the effect obtained for bubbles is verified by solving the full set of Newton's equations of motion for the particles near the critical surface determined from the simplified condition.

\section*{Acknowledgments}

The funding support received of the Network-35 of the Office of external
Activities (OEA) of the ICTP is greatly appreciated. In addition, both authors
would like to thank helpful conversations with Drs.  E. Altshuler and A. Lage.

\end{document}